\documentclass[12pt]{article}
\usepackage{amssymb}

\newcommand{\be}{\begin{equation}}
\newcommand{\ee}{\end{equation}}
\newcommand{\bea}{\begin{eqnarray}}
\newcommand{\eea}{\end{eqnarray}}
\newcommand{\barray}{\begin{array}}
\newcommand{\earray}{\end{array}}

\newcommand{\nn}{\nonumber}
\newcommand{\bitem}{\begin{itemize}}
\newcommand{\eitem}{\end{itemize}}
\newtheorem{teo}{Theorem}[section]
\newcommand{\bt}{\begin{teo}}
\newcommand{\et}{\end{teo}}
\newtheorem{Def}{Definition}[section]
\newcommand{\bd}{\begin{Def}}
\newcommand{\ed}{\end{Def}}
\newtheorem{lem}{Lemma}[section]
\newcommand{\bl}{\begin{lem}}
\newcommand{\el}{\end{lem}}
\newtheorem{prop}{Proposition}[section]
\newcommand{\bp}{\begin{prop}}
\newcommand{\ep}{\end{prop}}
\newtheorem{cor}{Corollary}[section]
\newcommand{\bc}{\begin{cor}}
\newcommand{\ec}{\end{cor}}
\newtheorem{ex}{Example}[section]
\newcommand{\bex}{\begin{ex}}
\newcommand{\eex}{\end{ex}}
\newtheorem{rem}{Remark}[section]
\newcommand{\br}{\begin{rem}}
\newcommand{\er}{\end{rem}}

\begin{document}

\begin{center}
{\Large \textbf{On consistency of determinants \\
on cubic lattices\footnotetext[1]{The work was completed with the
financial support of the Russian Foundation for Basic Research
(grant no. 08-01-00054) and the Programme for Support of Leading
Scientific Schools (grant no. NSh-1824.2008.1).}}}
\end{center}

\begin{center}
{\large \bf {O. I. Mokhov}}
\end{center}

\smallskip

Consider the square lattice $\mathbb{Z}^2$ with vertices at points
with integer-valued coordinates in $\mathbb{R}^2 = \{ (x_1, x_2)|\
x_k \in \mathbb{R},\ k =1, 2 \}$ and complex (or real) scalar fields
$u$ on the lattice $\mathbb{Z}^2$, $u : \mathbb{Z}^2 \rightarrow
\mathbb{C}$, that are defined by their values $u_{i_1 i_2}$, $u_{i_1
i_2} \in \mathbb{C}$, at each vertex of the lattice with the
coordinates $(i_1, i_2)$, $i_k \in \mathbb{Z}$, $k = 1, 2$. Consider
a class of two-dimensional discrete equations on the lattice
$\mathbb{Z}^2$ for the field $u$ that are defined by functions $Q
(x_1, x_2, x_3, x_4)$ of four variables with the help of the
relations \be Q (u_{i j}, u_{i + 1, j}, u_{i, j + 1}, u_{i + 1, j +
1}) = 0, \ \ \ \ i, j \in \mathbb{Z}, \label{1}\ee so that in each
{\it elementary $2 \times 2$ square of the lattice $\mathbb{Z}^2$},
that is, in each set of vertices of the lattice with coordinates of
the form $(i, j)$, $(i + 1, j)$, $(i, j + 1)$, $(i + 1, j + 1)$, $i,
j \in \mathbb{Z}$, the value of the field $u$ at one of vertices of
the square is defined by the values of the field at three other
vertices. In this case the field $u$ on the lattice $\mathbb{Z}^2$
is completely determined by fixing initial data, for example, on the
axes of coordinates of the lattice, $u_{i \, 0}$ and $u_{ 0 j}$, $i,
j \in \mathbb{Z}$. A particularly important role is played by {\it
integrable nonlinear discrete equations}. In [1]--[3] integrable
discrete equations of the form (\ref{1}) are singled out by the very
natural condition of {\it consistency on cubic lattices} (see also
[4]--[9]). Consider the cubic lattice $\mathbb{Z}^3$ with vertices
at points with integer-valued coordinates in $\mathbb{R}^3 = \{
(x_1, x_2, x_3)|\ x_k \in \mathbb{R},\  k =1, 2, 3 \}$ and fix
initial data $u_{i\, 0 0}$, $u_{ 0 j 0}$ and $u_{ 0 0 k}$, $i, j, k
\in \mathbb{Z}$, on the axes of coordinates of the cubic lattice. A
two-dimensional discrete equation (\ref{1}) is called {\it
consistent on the cubic lattice} if for initial data in general
position the discrete equation (\ref{1}) can be imposed in a
consistent way on all two-dimensional sublattices of the cubic
lattice $\mathbb{Z}^3$ at once (see [1]--[5]). Classifications of
discrete equations of the form (\ref{1}) that are consistent on the
cubic lattice have been studied in [4] and [8] under some additional
restrictions, see also [9]. The equation defined by determinants of
$2 \times 2$ matrices of values of the field at vertices of
elementary $2 \times 2$ squares of the lattice $\mathbb{Z}^2$: \be
u_{i, j + 1} u_{i + 1, j} - u_{i + 1, j + 1} u_{i j} = 0, \ \ \ \ i,
j \in \mathbb{Z}, \label{2}\ee is an example of such two-dimensional
nonlinear integrable discrete equation consistent on the cubic
lattice. The equation (\ref{2}) is linear with respect to each
variable and invariant with respect to the full symmetry group of
square. The fixing of arbitrary nonzero initial data $u_{i \, 0}$
and $u_{0 j}$, $i, j \in \mathbb{Z}$, on the axes of coordinates of
the lattice $\mathbb{Z}^2$ completely determines the field $u$ on
the lattice $\mathbb{Z}^2$ satisfying the discrete equation
(\ref{2}), and the fixing of arbitrary nonzero initial data $u_{i \,
0 0}$, $u_{0 j 0}$ and $u_{0 0 k}$, $i, j, k \in \mathbb{Z}$, on the
axes of coordinates of the lattice $\mathbb{Z}^3$ completely
determines the field $u$ on the lattice $\mathbb{Z}^3$ satisfying
the discrete equation (\ref{2}) on all two-dimensional sublattices
of the cubic lattice $\mathbb{Z}^3$. The integrability (in the broad
sense of the word) of the discrete equation (\ref{2}) is obvious
since it can be easily linearized: $ \ln u_{i, j + 1} + \ln u_{i +
1, j} - \ln u_{i + 1, j + 1} - \ln u_{i j} = 0, \ i, j \in
\mathbb{Z}.$ In this work, we consider the question on consistency
on cubic lattices for discrete nonlinear equations defined by
determinants of matrices of higher orders (for orders $N > 2$). The
condition of consistency on cubic lattices in the form as it was
defined above is not satisfied for these discrete equations if $N >
2$. We prove that an other, modified, condition of consistency on
cubic lattices, which is proposed in this work, is satisfied for
determinants of matrices of arbitrary orders.

Consider a discrete equation on the lattice $\mathbb{Z}^2$ defined
by a relation for values of the field $u$ at vertices of the lattice
$\mathbb{Z}^2$ that form {\it elementary $3 \times 3$ squares}: \be
Q (u_{i j}, \ldots, u_{i + s, j + r}, \ldots, u_{i + 2, j + 2}) = 0,
\ \  0 \leq s, r \leq 2, \ \ i, j \in \mathbb{Z}. \label{4}\ee The
fixing of initial data $u_{i \, 0}$, $u_{i 1}$, $u_{0 j}$ and $u_{1
j}$, $i, j \in \mathbb{Z}$, completely determines the field $u$ on
the lattice $\mathbb{Z}^2$ satisfying the discrete equation
(\ref{4}). Consider the cubic lattice $\mathbb{Z}^3$ and the
condition of consistency on all two-dimensional sublattices of the
cubic lattice $\mathbb{Z}^3$ for the discrete equation (\ref{4}).
Initial data can be specified, for example, at the following
vertices of the lattice: $u_{i\, 0 0}$, $u_{i 1 0}$, $u_{i\, 0 1}$,
$u_{i 1 1}$, $u_{0 j 0}$, $u_{1 j 0}$, $u_{0 j 1}$, $u_{1 j 1}$,
$u_{0 0 k}$, $u_{1 0 k}$, $u_{0 1 k}$ and $u_{1 1 k}$, $i, j, k \in
\mathbb{Z}$. In the cube $\{(i, j, k), \ 0 \leq i, j, k \leq 2 \}$,
the values $u_{2 0 2}$, $u_{2 1 2}$, $u_{2 2 0}$, $u_{2 2 1}$, $u_{0
2 2}$ and $u_{1 2 2}$ are determined by the relations (\ref{4}), and
three relations must be satisfied for the value $u_{2 2 2}$ on three
faces of the cube at once. Consider the discrete nonlinear equation
on the lattice $\mathbb{Z}^2$ defined by determinants of the
matrices of values of the field $u$ at vertices of the lattice
$\mathbb{Z}^2$ that form elementary $3 \times 3$ squares: \bea &&
u_{i, j + 2} u_{i + 1, j + 1} u_{i + 2, j} + u_{i, j + 1} u_{i + 1,
j} u_{i + 2, j + 2} + u_{i, j} u_{i + 1, j +
2} u_{i + 2, j + 1} - \nn \\
&& - u_{i, j} u_{i + 1, j + 1} u_{i + 2, j + 2} - u_{i, j + 2} u_{i
+ 1, j} u_{i + 2, j + 1} - u_{i, j + 1} u_{i + 1, j + 2} u_{i + 2,
j} = 0,  \label{5}\eea where $i, j \in \mathbb{Z}$. The equation
(\ref{5}) is linear with respect to each variable and invariant with
respect to the full symmetry group of the configuration of vertices
of the lattice $\mathbb{Z}^2$ that form elementary $3 \times 3$
squares. For initial data in general position, the discrete equation
(\ref{5}) is not consistent on two-dimensional sublattices of the
cubic lattice $\mathbb{Z}^3$.

Consider an other, modified, condition of consistency on cubic
lattices for the discrete nonlinear equation (\ref{5}). We shall
require that the discrete equation (\ref{5}) would be satisfied not
only on all two-dimensional sublattices of the cubic lattice
$\mathbb{Z}^3$, but also on all unions of any two intersecting
two-dimensional sublattices of the cubic lattice $\mathbb{Z}^3$,
that is, the corresponding elementary $3 \times 3$ squares, on which
the discrete equation (\ref{5}) is considered, can be bent at right
angle passing from one of two-dimensional sublattices to another,
for example, $\{(i, 0, 0), (i, 1, 0), (i, 0, 1), i = 0, 1, 2\}$,
$\{(0, j, 0), (1, j, 0), (0, j, 1), j = 0, 1, 2\}$ and $\{(0, 0, k),
(1, 0, k), (0, 1, k), k = 0, 1, 2\}$. In this case, initial data can
be specified, for example, at the following vertices of the cubic
lattice $\mathbb{Z}^3$: $u_{i\, 0 0}$, $u_{i\, 0 1}$, $u_{2 j 0}$,
$u_{ 2 j 1}$, $u_{ 1 0 k}$, $u_{ 2 0 k}$, $u_{ 1 1 0}$ and $u_{1 1
2}$, $i, j, k \in \mathbb{Z}$. We shall also call the corresponding
discrete equations {\it consistent on cubic lattices}. Then the
following theorem holds.

{\bf Theorem 1}. {\it For arbitrary initial data in general position
the nonlinear discrete equation {\rm (\ref{5})} can be satisfied in
a consistent way on all unions of any two two-dimensional
sublattices of the cubic lattice $\mathbb{Z}^3$, that is, the
discrete nonlinear equation {\rm (\ref{5})} is consistent on the
cubic lattice $\mathbb{Z}^3$.}

A similar property of consistency on cubic lattices holds for
determinants of arbitrary orders $N$.

{\bf Theorem 2}. {\it For any given integer $N$, the discrete
nonlinear equation defined by determinants of the matrices of values
of the field $u$ at vertices of the lattice $\mathbb{Z}^2$ that form
elementary $N \times N$ squares is consistent on the cubic lattice
$\mathbb{Z}^3$.}

\smallskip

\smallskip

\begin{center}
{\bf Bibliography}
\end{center}

\smallskip

\noindent [1] F.W. Nijhoff, A.J. Walker, {\it Glasgow Math. J.} {\bf
43A} (2001), 109--123.

\noindent [2] F.W. Nijhoff, {\it Phys. Lett. A} {\bf 297} (2002),
49--58.

\noindent [3] A.I. Bobenko, Yu.B. Suris, {\it Int. Math. Res.
Notices} {\bf 11} (2002), 573--611.

\noindent [4] V.E. Adler, A.I. Bobenko, Yu.B. Suris, {\it Comm.
Math. Phys.} {\bf 233} (2003), 513--543.

\noindent [5] A.I. Bobenko, Yu.B. Suris, arXiv: math/0504358.

\noindent [6] A.I. Bobenko, Yu.B. Suris, {\it Uspekhi Mat. Nauk}
{\bf 62}:1 (2007), 3--50; English transl., A.I. Bobenko, Yu.B.
Suris, {\it Russian Math. Surveys} {\bf 62}:1 (2007), 1–-43.

\noindent [7] A.P. Veselov, {\it Uspekhi Mat. Nauk} {\bf 46}:5
(1991), 3--45. English transl., A.P. Veselov, {\it Russian Math.
Surveys} {\bf 46}:5 (1991), 1–-51.

\noindent [8] V.E. Adler, A.I. Bobenko, Yu.B. Suris, arXiv:
abs/0705.1663.

\noindent [9] S.P. Tsarev, T. Wolf, {\it Lett. Math. Phys.} {\bf
84}:1 (2008), 31--39.

\begin{flushleft}
Centre for Non-Linear Studies,\\
L.D.Landau Institute for Theoretical Physics,\\
Russian Academy of Sciences;\\
Department of Geometry and Topology,\\
Faculty of Mechanics and Mathematics,\\
M.V.Lomonosov Moscow State University;\\
{\it E-mail\,}: mokhov@mi.ras.ru; mokhov@landau.ac.ru; mokhov@bk.ru\\
\end{flushleft}
\end{document}